\documentclass[letterpaper, aps, prd, twocolumn, superscriptaddress, showpacs, nofootinbib]{revtex4}
\pdfoutput=1

\usepackage{graphicx}
\usepackage{bm}
\usepackage{dcolumn}
\usepackage{color}
\usepackage[sort&compress]{natbib}
\usepackage[latin9]{inputenc}
\usepackage{url}
\usepackage{subfigure}
\usepackage{float}
\usepackage{longtable}
\usepackage{amssymb}
\usepackage{amsmath}
\usepackage{amsfonts}
\usepackage{array}
\usepackage{graphicx}
\usepackage{multirow}
\usepackage{xcolor}
\usepackage{ulem}
\usepackage{hyperref}
\newcommand\MyBox[2]{
  \fbox{\lower0.75cm
    \vbox to 1.2cm{\vfil
      \hbox to 1.2cm{\hfil\parbox{1.0cm}{#1\\#2}\hfil}
      \vfil}%
  }%
}

\begin{document}

%%%%%%%%%%%%%%%%%%%%%%%%%%%%%%%%%%%%%%%%%%%%%%%%%%

%%%%%%%%%%%%%%%%%%% TITLE PAGE %%%%%%%%%%%%%%%%%%%

\title{Reconstructing Core-Collapse Supernova Gravitational-Wave Signals with Transdimensional Bayesian Inference}

\newcommand*{\swin}{Centre for Astrophysics and Supercomputing, Swinburne University of Technology, Hawthorn, VIC 3122, Australia.}
\affiliation{\swin}

\newcommand*{\ozgrav}{ARC Centre of Excellence for Gravitational Wave Discovery (OzGrav), Melbourne, Australia.}
\affiliation{\ozgrav}

\newcommand*{\monash}{School of Physics and Astronomy, Monash University, Vic 3800, Australia.}
\affiliation{\monash}

\newcommand*{\unimelb}{School of Physics, University of Melbourne, Parkville, Vic 3010, Australia. }
\affiliation{\unimelb}

% The list of authors, and the short list which is used in the headers.
% If you need two or more lines of authors, add an extra line using \newauthor
\author{Hayden Chapman}  \affiliation{\swin} 
\author{Jade Powell}  \email{dr.jade.powell@gmail.com} \affiliation{\swin} \affiliation{\ozgrav} 
\author{Nir Guttman} \affiliation{\monash} \affiliation{\ozgrav}
\author{Yi Shuen Christine Lee} \affiliation{\unimelb} \affiliation{\ozgrav}
\author{Paul D. Lasky} \affiliation{\monash} \affiliation{\ozgrav}

% These dates will be filled out by the publisher
% \date{Accepted XXX. Received YYY; in original form ZZZ}

% Prints the current year, for the copyright statements etc. To achieve a fixed year, replace the expression with a number. 

\label{firstpage}

\begin{abstract}
Core-collapse supernovae (CCSNe) are  promising future sources of gravitational waves for current and next-generation observatories. Reconstructing CCSN gravitational-wave signals is challenging as they contain stochastic elements, have multiple complex features, and cover a wide frequency band. The stochasticity of the signal in particular motivates the need for morphology-independent reconstruction techniques which, once observed, will enable us to infer properties of the newly born proto-neutron star, the rotation, and the unknown CCSN explosion mechanism. In this work, we investigate the reconstruction of gravitational-wave signals from CCSNe using the transdimensional Bayesian inference framework \texttt{tBilby}. We demonstrate the method using simulated signals in synthetic Advanced LIGO detector noise at a range of signal-to-noise ratios. We reconstruct the signals using two types of wavelets: sine Gaussians and chirplets. We calculate overlaps between injected and reconstructed signals of up to 85\%. We find that even when reconstruction overlap values are low, enough of the time-frequency structure of the dominant mode is captured to still make statements about the size of the evolving proto-neutron star. These capabilities establish \texttt{tBilby} as a powerful tool for gravitational-wave astronomy with burst sources. 
\end{abstract}

\maketitle

%%%%%%%%%%%%%%%%%%%%%%%%%%%%%%%%%%%%%%%%%%%%
\section{Introduction}
\label{sec:intro}

Stars with a zero-age main sequence (ZAMS) mass $\text{M} \gtrsim 8\,\text{M}_{\odot}$ end their lives in a core-collapse supernova (CCSN). Once the iron core of the star reaches its effective Chandrasekhar mass, runaway collapse ensues due to gravitational instability. Once the core reaches nuclear densities, the supersonically infalling inner core rebounds, causing an outward shock. This outward shock then begins to slow down due to heating up the subsonically infalling outer core. A few hundred milliseconds after collapse, the shock front stalls at a radius of a few hundred km. How the shockwave then regains energy, the CCSN explosion mechanism, is not fully understood. For a detailed review of CCSN explosion dynamics, see~\cite{burrows_21, jerkstrand_26}.   

There are multiple possible pathways to shock revival.
The explosions of non or slowly rotating progenitor stars are believed to be neutrino driven \cite{janka_17}. The neutrino heating mechanism occurs due to the immense amount of neutrinos being emitted from the core transferring a small amount of their energy into the gain region of the stalled shock front \cite{janka_17}. Rapidly rotating progenitors with strong magnetic fields, which are expected to only be~$\approx1\%$ of CCSN events, may explode by the magneto-rotational explosion mechanism \cite{kuroda_20}. While these mechanisms provide possible solutions to the necessary shock revival, there is no empirical evidence of either pathway from the constraints placed by the thousands of detailed, multi-wavelength electromagnetic observations. 
 
To understand the mechanism responsible for powering CCSNe, gravitational-wave or neutrino observations are needed, as these messengers are emitted directly from the core of the explosion. 
The current gravitational-wave observatories LIGO \cite{ligo_15}, Virgo \cite{virgo_15} and KAGRA \cite{kagra_19}, have been successful at detecting gravitational-wave signals from merging black holes and neutron stars~\cite{gwtc-5}. As the detectors improve their sensitivity, other lower amplitude sources will begin to be observed \cite{powell_25}, with one of the most promising future sources being CCSNe. Searches for gravitational-wave emission from CCSNe have been performed but so-far without a detection \cite{szczepanczyk_24, sn2023ixf_25}.  
The current network of gravitational-wave detectors are expected to be sensitive to CCSNe in our own Galaxy, or the Magellanic Clouds, which are thought to occur only once or twice per century~\cite{szczepanczyk_21}. 

Gravitational-wave emission from CCSNe is predicted through hydrodynamical simulations \cite[e.g., see][]{mueller_24}. Although there are still many uncertainties in CCSN physics, common features in the gravitational-wave emission have been observed between different simulation codes. The main gravitational-wave emission feature is a dominant proto-neutron star (PNS) mode that starts at a few hundred Hz and increases in frequency with time up to a few thousand Hz \cite{kuroda_17, oconnor_18, powell_19, andresen_19, radice_19, vartanyan_20, pan_21, jakobus_23, mezzacappa_24}. The gravitational-wave frequency of this mode increases as the PNS gains mass and shrinks in radius. Other, lower amplitude, PNS modes can also be present in the gravitational-wave signal \cite{burrows_26}. Additionally, there may be a lower frequency mode due to the standing accretion shock instability (SASI) \cite{blondin_03, walk_23, foglizzo_24}. The SASI mode mode starts below 200\,Hz and slowly increases in frequency until either the shock is revived or the star collapses to a black hole. In the first 100\,ms after the core bounce, there may also be a short burst of high amplitude gravitational-wave emission around 100\,Hz due to prompt convection \cite{bruenn_94, cusinato_26}. 

Current search algorithms for gravitational-wave emission from CCSNe are model agnostic, and make minimal assumptions about the source~\cite[e.g.,][]{szczepanczyk_21}. After a CCSN has been detected, we will need to reconstruct the gravitational-wave signal and attempt to relate the properties of the gravitational-wave emission to the astrophysical properties of the source. The two main algorithms currently used to reconstruct gravitational-wave burst signals, without making any assumptions about the signal morphology, are Coherent WaveBurst~\cite[cWB;][]{drago_21} and BayesWave \cite{cornish_15}. BayesWave uses sums of wavelets and Bayesian reverse jump MCMC to reconstruct gravitational-wave signals. The authors in \cite{raza_22}, showed how well the BayesWave algorithm can reconstruct CCSN signals in Advanced LIGO noise for a range of different signal-to-noise ratios (SNRs). The authors in \cite{lee_25}, use both BayesWave and cWB to reconstruct CCSN signals in different frequency bands. However, much work is still needed to improve CCSN waveform reconstructions, as these previous works have shown that only $\sim 60$\% of some CCSN signals are being reconstructed even at high (20-30) SNR values~\cite{szczepanczyk_21, raza_22, lee_25}. Further work is also needed to understand how the waveform reconstruction accuracies relate to what we can learn about the astrophysics of the source. 

Other works have used more modelled approaches to determine the astrophysical parameters of CCSN gravitational-wave sources. The core bounce in rotating models results in a time series spike that is understood well enough to develop modelled parameter estimation methods~\cite{edwards_21, pastor_24}. Universal relations \cite{torres_19, sotani_21}, which describe the relationship between the gravitational-wave frequency and the size of the PNS, have also been used to infer PNS parameters~\cite{powell_22, bruel_23}. Principal component analysis has also been applied to waveforms from numerical simulations to enable model selection studies to determine the CCSN explosion mechanism~\cite{powell_24}. 

In this work, we perform unmodelled reconstructions of CCSN signals using the transdimensional Bayesian inference framework~\texttt{tBilby}~\cite{tong_25}, which is built as a wrapper for the Bayesian inference library~\texttt{Bilby}~\cite{bilby_paper, bilby_second_paper}. The \texttt{tBilby} code has been used previously to reconstruct gravitational-wave signals from binary black holes~\cite{tong_25}, but this is the first time that this method has been applied to CCSN signals. We use four different CCSN models added to LIGO design sensitivity noise. As in previous work, we measure the overlap between the reconstructed and injected waveforms. 
We go beyond the previous unmodelled reconstruction studies by using multiple types of wavelets to determine which perform the best for CCSN waveforms, and by investigating what astrophysics can be inferred from reconstructions with different overlap values. As some CCSN waveform features are stochastic, capturing 100\% of the emission may not be necessary for the correct inference of the astrophysical source parameters. 

This paper is structured as follows: In Section \ref{sec:supernova}, we discuss the CCSN waveforms that are used in this work. In Section \ref{sec:tbilby}, we describe the transdimensional inference algorithm and the setup of our analysis. In Section \ref{sec:results}, we show our overlap values and example reconstructions. The conclusions are given in Section \ref{sec:conclusions}.

%%%%%%%%%%%%%%%%%%%%%%%%%%%%%%%%%%%%%%%%%%%%%%%%
\section{Core-collapse Supernovae}
\label{sec:supernova}

\begin{figure}
\includegraphics[width=\columnwidth]{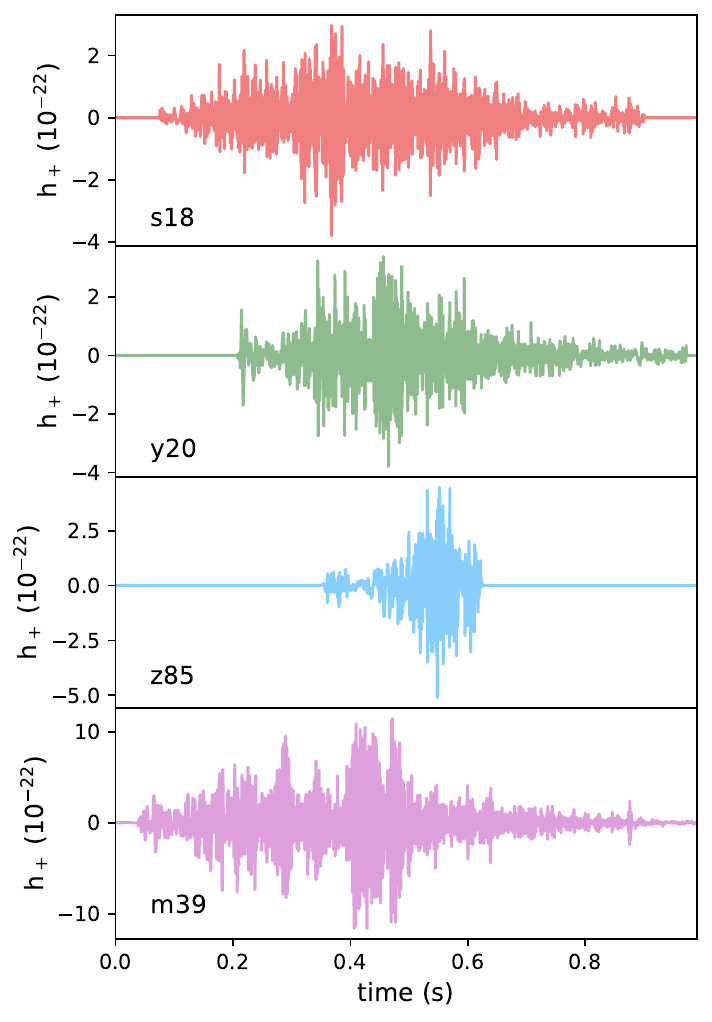}
\caption{The CCSN simulated signals used in this study at a distance of 10\,kpc. The s18, y20 and z85 models are non-rotating. Model z85 is short duration as it rapidly forms a black hole due to its high mass. The m39 model has higher gravitational-wave amplitudes as it is rapidly rotating.
}
\label{fig:waveforms}
\end{figure}

We use four different CCSN waveforms from numerical simulations, shown in Figure 
\ref{fig:waveforms}, to represent the gravitational-wave signal we might expect to see from a real CCSN detection. For all waveforms, we use a sample rate of 2048\,Hz, and a lower frequency cut off of 30\,Hz. We pick this relatively low sample rate for a faster analysis, but as some waveforms from numerical simulations reach higher frequencies, a higher maximum frequency would be needed for the analysis of a real CCSN detection. 

The first is model s18, from \cite{powell_19}, which has a ZAMS mass of $18\,\mathrm{M}_{\odot}$ and is non-rotating. This model is dominated by the high frequency f/g-mode emission for about 0.6\,s which reaches a maximum frequency $\sim 950$\,Hz.

The second waveform is model y20 from \cite{powell_20}, which has a ZAMS mass of $20\,\mathrm{M}_{\odot}$ and is non-rotating. This waveform has some gravitational waves from prompt convection in the first 0.1\,s, after which the signal is dominated by the f/g-mode for the next $\sim0.5$\,s. 

The third waveform is model m39 from \cite{powell_20}. It has a ZAMS mass of $39\,\mathrm{M}_{\odot}$, and is different to the other models considered because it is rapidly rotating. The gravitational-wave emission is still dominated by the f/g-mode, however the amplitude is significantly larger, and so it can be detected from much larger distances than the other models considered here. 

The fourth model is the z85 model from \cite{powell_21}, with the LS220 equation of state. It has a very high ZAMS mass of $85\,M_{\odot}$, which results in very rapid formation of a black hole, so the gravitational-wave signal is very short at less than 0.3\,s. Shock revival still occurred before black hole formation.

%%%%%%%%%%%%%%%%%%%%%%%%%%%%%%%%%%%%%%%%%%%%%
\section{Inference Method}
\label{sec:tbilby}

\begin{figure}
\includegraphics[width=\columnwidth]{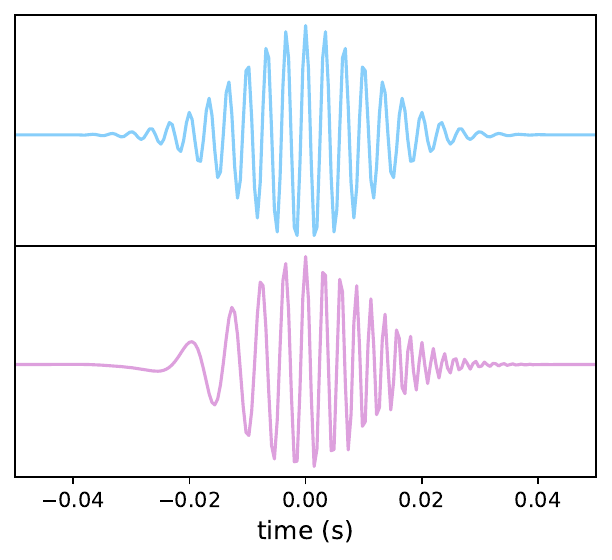}
\caption{ 
Examples of the wavelets used in this study. The top panel shows a sine Gaussian wavelet. The bottom panel shows a chirplet wavelet. The wavelets were calculated with a Q of 30, and a frequency of 300\,Hz. Chirplets differ from sine Gaussians only by their rate of change of frequency with time. 
}
\label{fig:chirp}
\end{figure}

The \texttt{tBilby} code~\cite{tong_25} implements transdimensional Bayesian inference in which the number of paramaters $N$ is itself a free parameter. In other words, we use Bayes' theorem to reconstruct the posterior distribution $p({\boldsymbol \theta}|{\bf d})$ for data $d$ and model parameters $\boldsymbol{\theta}$, such that 
\begin{equation}
    \boldsymbol{\theta}=\left\{\theta_1,\theta_2,\ldots,\theta_N,N\right\}.
\end{equation}
There are a number of different possible implementations of transdimensional sampling, including algorithms such as reversible jump MCMC. The \texttt{tBilby} software instead treats $N$ similarly to any other discrete free parameter. Using a nested sampling algorithm, at each step the sampler draws a sample from $N$ as well as values from all other $\boldsymbol{\theta}$. All $\theta_k>\theta_N$ for that particular draw are treated as ghost parameters, in the sense that they are not used in the likelihood evaluation. The final posterior distribution is achieved by marginalising over all ghost parameters---for details, see~\cite{tong_25}.

In this study, we use two different sets of basis functions within \texttt{tBilby} to reconstruct our signal. These are sine-Gaussian wavelets and chirplets. Chirplets are similar to sine Gaussian wavelets, except their instantaneous frequency is free to change with time. An example sine Gaussian is shown in the top panel of Figure \ref{fig:chirp}, and an example chirplet in the bottom panel. 

A sine-Gaussian in the frequency domain is given by,
\begin{equation}
\begin{aligned}
    \Psi(f;A, f_0, Q, t_0, \phi_0) = \frac{A\sqrt{\pi}\tau}{2}\text{exp}[-\pi^2\tau^2\Delta f^2](\text{exp}[i\phi_0] \\ 
    + \text{exp}[-i\phi_0]\text{exp}[-\frac{Q^2f}{f_0}]),
\end{aligned}
\end{equation}
where $Q$ is the quality factor, $f$ is the frequency array, $f_{0}$ is the central frequency, $\Delta f = f - f_0$, $\tau$ is the duration, $A$ is the amplitude, $t_{o}$ is the central time, and $\phi_{0}$ is the phase. 
Similarly, the chirplet model in the frequency domain is defined as, 
\begin{equation}
\begin{aligned}
    \Psi(f;A, f_0, \dot{f_0}, Q, t_0, \phi_0) = \frac{A\sqrt{\pi}\tau}{2(1 + \pi^2\beta^2)^{1/4}} \text{exp}[-\frac{\pi^2\tau^2\Delta f^2}{1 + \pi^2\beta^2}] \\ \text{exp}[-2\pi ift_0](\text{exp}[\frac{i(\phi_0 + \delta-\pi^3\beta\tau^2\Delta f^2)}{1 + \pi^2\beta^2}] + \\
    \text{exp}[-\frac{Q^2f}{f_0}]\text{exp}[-\frac{i(\phi_0 + \delta-\pi^3\beta^2\tau^2\Delta f^2)}{1 + \pi^2\beta^2}]),
\end{aligned}
\end{equation}
where an overdot denotes a derivative with respect to time, $\delta = \frac{1}{2}\text{arctan}(\pi\dot{f_0}\tau^2)$ and $\beta = \dot{f_0}\tau^2$. The difference between the two models is that the chirplet has an extra parameter $\dot{f_0}$ which allows the function to change in frequency with respect to time. When $\dot{f_0} = 0$, the chirplet reduces to a sine-Gaussian. 

For our analysis, we use the standard gravitational-wave transient likelihood function~\cite{veitch15}. We use uniform priors on all of our model parameters, and assume that the time and sky position of the gravitational-wave signal is already known. We use the same sky position for all signals. We use simulated Gaussian noise with an Advanced LIGO power-spectral density at design sensitivity \cite{ligo_noise_curve}, and inject all the signals into a two LIGO detector network of Hanford and Livingston. We vary the distance of the source to achieve SNR values between 20 and 60. Below this minimum value, it is difficult for the current CCSN search algorithms to find a signal \cite{szczepanczyk_21}. We set the maximum number of wavelets to 15. The dynesty sampler is used \cite{speagle_20}, with the maximum number of live points set to 1500.

%%%%%%%%%%%%%%%%%%%%%%%%%%%%%%%%%%%%%%%%%%%%%%%%
\section{Results}
\label{sec:results}

We calculate the overlap $\mathcal{O}$ between the injected waveforms and the reconstructed signals, which is given by 
\begin{equation}
\mathcal{O} = \frac{\langle \tilde{h}_{\rm inj} | \tilde{h}_{\rm rec} \rangle}{ \sqrt{ \langle \tilde{h}_{\rm inj} | \tilde{h}_{\rm inj} \rangle \langle \tilde{h}_{\rm rec} | \tilde{h}_{\rm rec} \rangle }}
\end{equation}
where $\tilde{h}_{\rm inj}$ is the Fourier-domain injected waveform, and $\tilde{h}_{\rm rec}$ is the recovered waveform. For a given noise power spectral density $S_n(f)$,  the noise weighted inner product is 
\begin{equation}
\langle \tilde{h}_{a} | \tilde{h}_{b} \rangle = 4 Re \int_{0}^{\infty} \frac{\tilde{h}_{a}(f) \tilde{h}_{b}^{*}(f) }{ S_{n}(f) } df. 
\label{eq:overlap}
\end{equation}
Overlaps go between zero and one, where the latter represents perfect signal recovery. 

\begin{figure*}
\includegraphics[width=\textwidth]{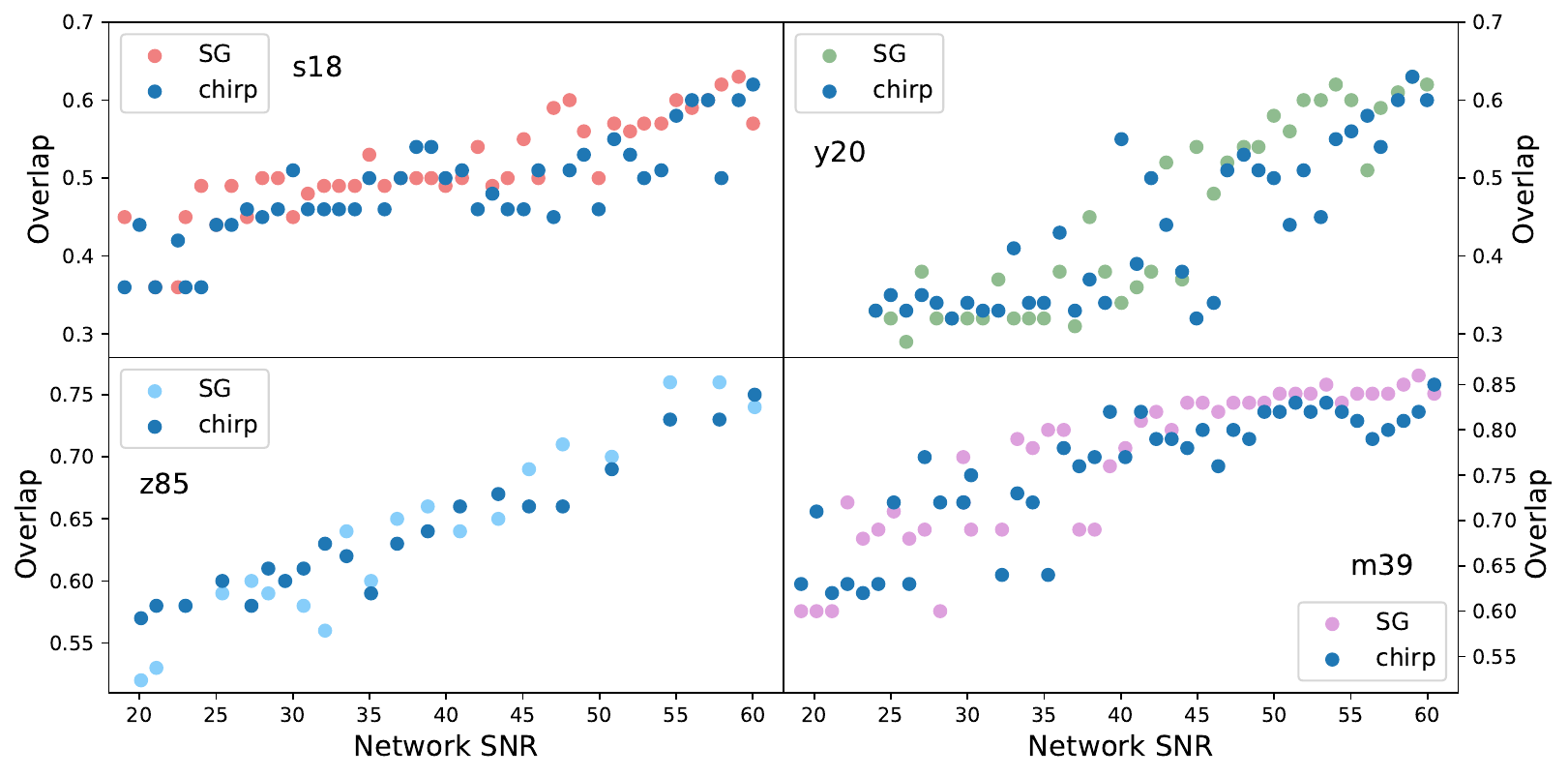}
\caption{Overlap between the injected and reconstructed waveforms, calculated using Equation \ref{eq:overlap}, as a function of the signal-to-noise ratio. ``SG" is the overlap calculated with the sine-Gaussian signal model. ``Chirp" is the overlap calculated with the chirplet signal model. Higher overlaps are generally obtained using the sine Gaussian signal model. The m39 model has the highest overlaps of all the models considered here. 
}
\label{fig:overlaps}
\end{figure*}

In Figure \ref{fig:overlaps} we show overlap as a function of network SNR for each of our injected signals, and colour-coded by the sine-Gaussian and chirplet reconstructions. The overlap shown is that of the median posterior value from the posterior distribution as calculated in \texttt{tBilby}. The maximum overlap is dependent on the different waveform models. For s18 models, the minimum waveform overlap is $\sim 0.35$ and the maximum is $\sim 0.65$. Similar results are found for the y20 model. The other two models reach higher overlap values. The z85 model has a minimum overlap of $\sim 0.5$ and a maximum of over $\sim 0.75$, whereas m39 has the best overlap reaching over $\sim 0.85$ at the highest SNRs considered. These overlap values are a little lower than those obtained in previous work with BayesWave~\cite{raza_22}. We observe no significant differences between the overlap values using the sine Gaussian or the chirplet wavelets. This is consistent with the results of previous work for reconstructing other types of gravitational-wave signals with chirplets \cite{millhouse_18}.

\begin{figure*}
\includegraphics[width=\textwidth]{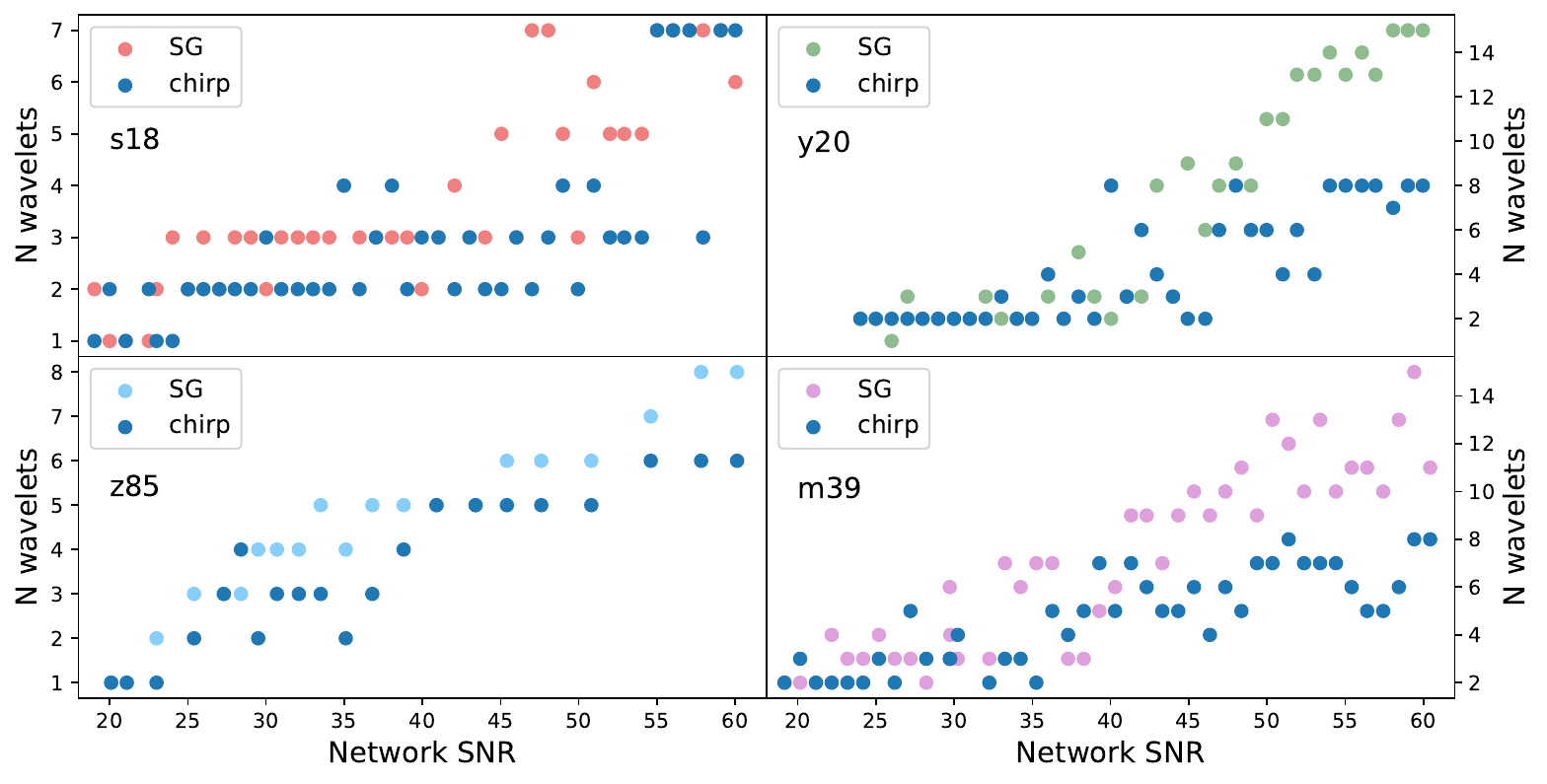}
\caption{For each injected waveform, we show the number of wavelets used to reconstruct the signal as a function of the signal-to-noise ratio. We use the maximum posterior values. Louder signals generally require more wavelets to reconstruct the models. A larger number of basis functions are generally required to capture the signal features with the sine Gaussian wavelets than with the chirplets.  
}
\label{fig:wavelets}
\end{figure*}

We show the number of wavelets picked by the sampler in Figure \ref{fig:wavelets}. We define the number of wavelets as the maximum likelihood value. The number of wavelets required to reconstruct the signal increases as the SNR increases and more of the signal features become visible above the noise level. Different CCSN models require a significantly different number of wavelets. The s18 and z85 models require the least wavelets, with only seven or eight sine Gaussian wavelets needed at even the highest SNR values considered here. The m39 and y20 models need up to 15 wavelets at the highest SNR values. 

\begin{figure*}
\includegraphics[width=\textwidth]{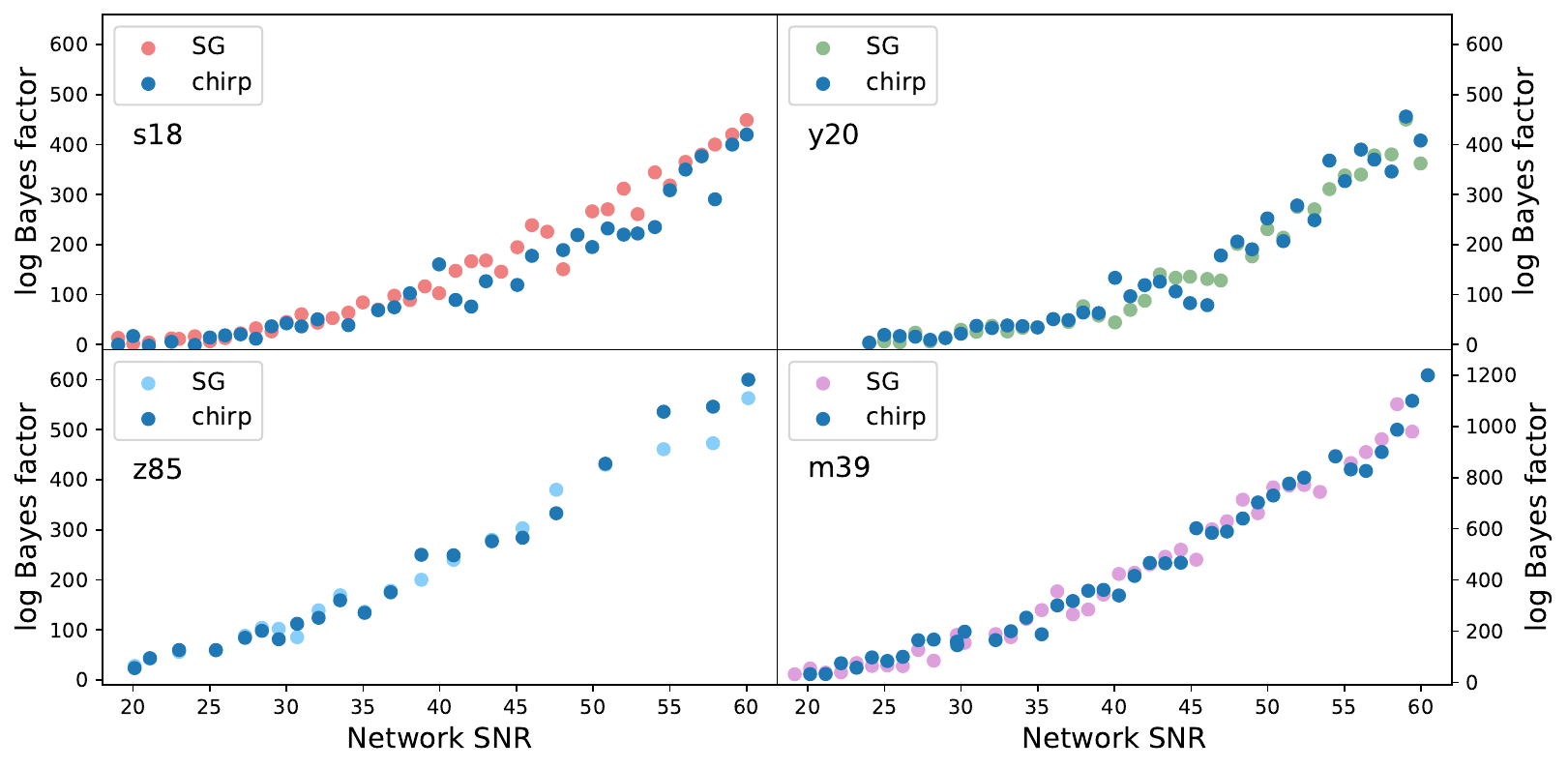}
\caption{
The log Bayes factors for models s18, y20, m39 and z85 at different signal-to-noise ratios. There are no significant differences between the chirplet and sine Gaussian wavelets.    
}
\label{fig:bayesfactor}
\end{figure*}

The sine Gaussian model needs a significantly higher number of wavelets to reconstruct the CCSN waveforms, even if there is not a significant increase in overlap. This result is also consistent with findings of similar previous work \cite{millhouse_18}. This means that the chirplet model may be better for a faster analysis, as the sampler is faster if only half as many wavelets are needed. We show the log Bayes factors in Figure \ref{fig:bayesfactor}. Similar Bayes factors are found between the sine Gaussian and chirplets, even with the large difference in number of wavelets. 

\begin{figure*}
\includegraphics[width=\textwidth]{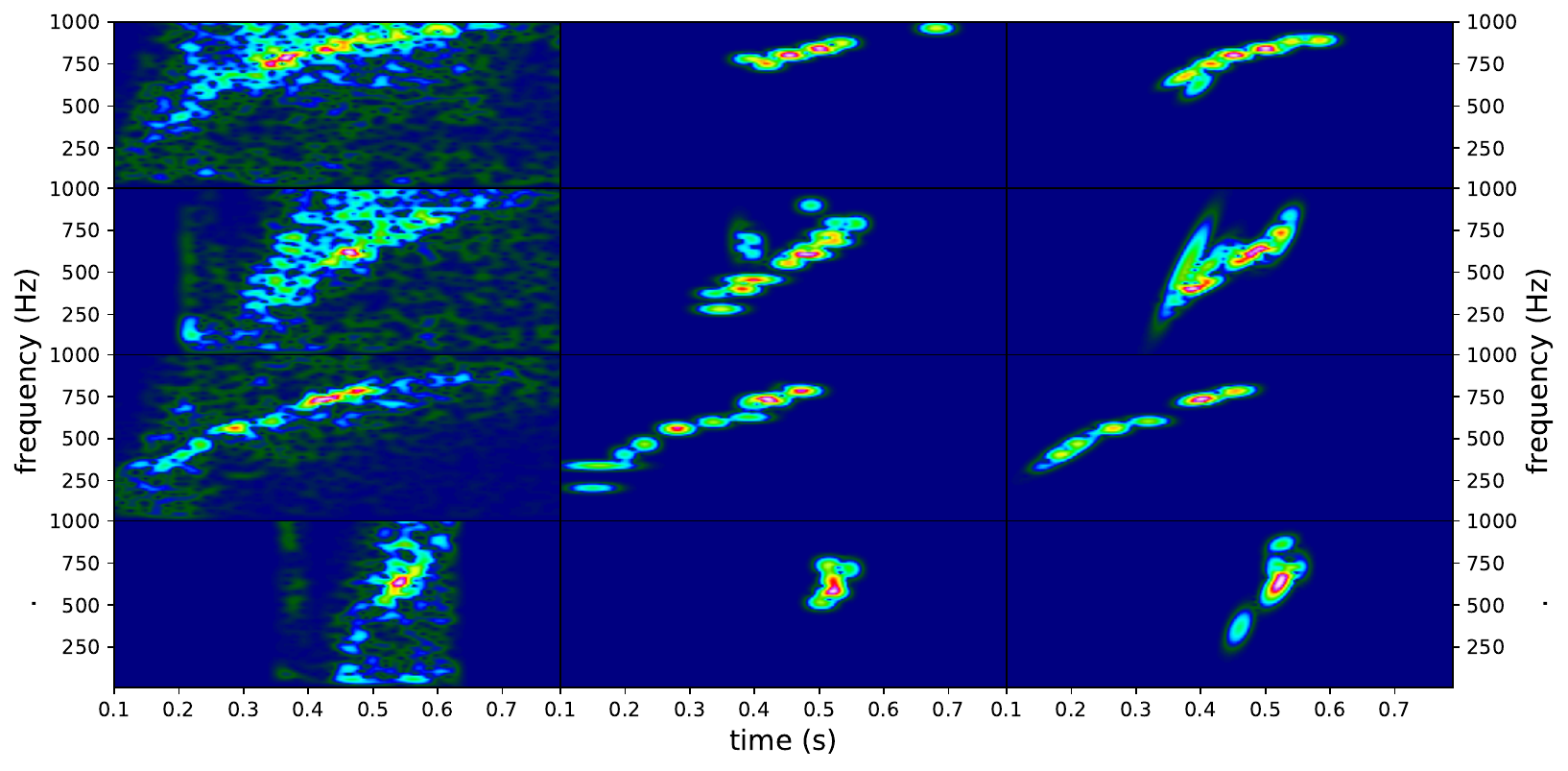}
\caption{
Gravitational-wave spectrograms. From top to bottom are models s18, y20, m39 and z85. The left column shows a spectrogram of the injected waveform. The middle column shows the reconstructed waveform for a signal injected with a signal-to-noise ratio of 60 using the sine Gaussian  wavelets. The right column shows the same for signals reconstructed using the chirplet wavelets.  
}
\label{fig:recons_snr60}
\end{figure*}

\begin{figure*}
\includegraphics[width=\textwidth]{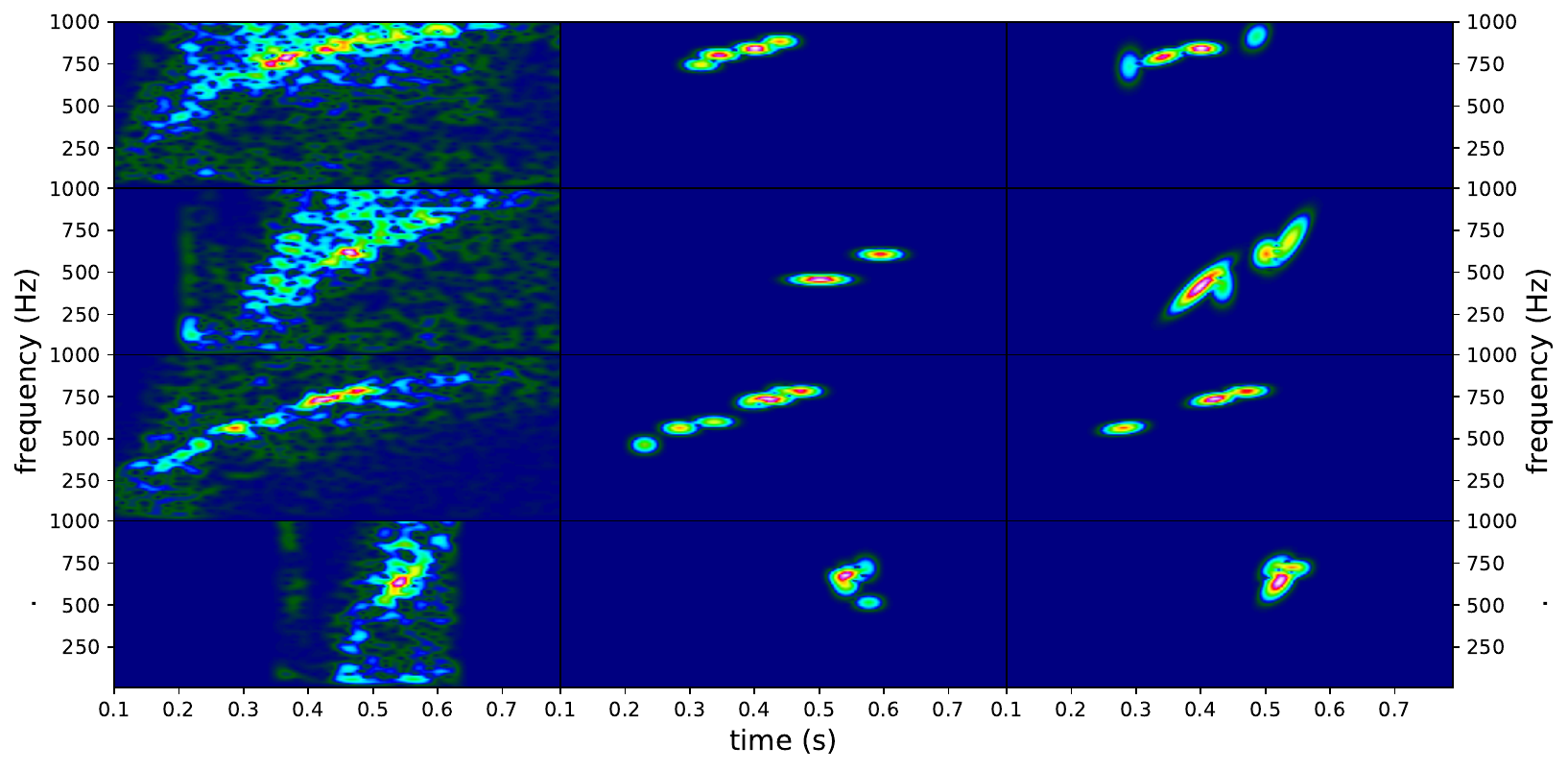}
\caption{ 
Same as Figure~\ref{fig:recons_snr60}, however the signal-to-noise of the injected signal is 35. A significant part of the mode is still visible. 
}
\label{fig:recons_snr35}
\end{figure*}

\begin{figure*}
\includegraphics[width=\textwidth]{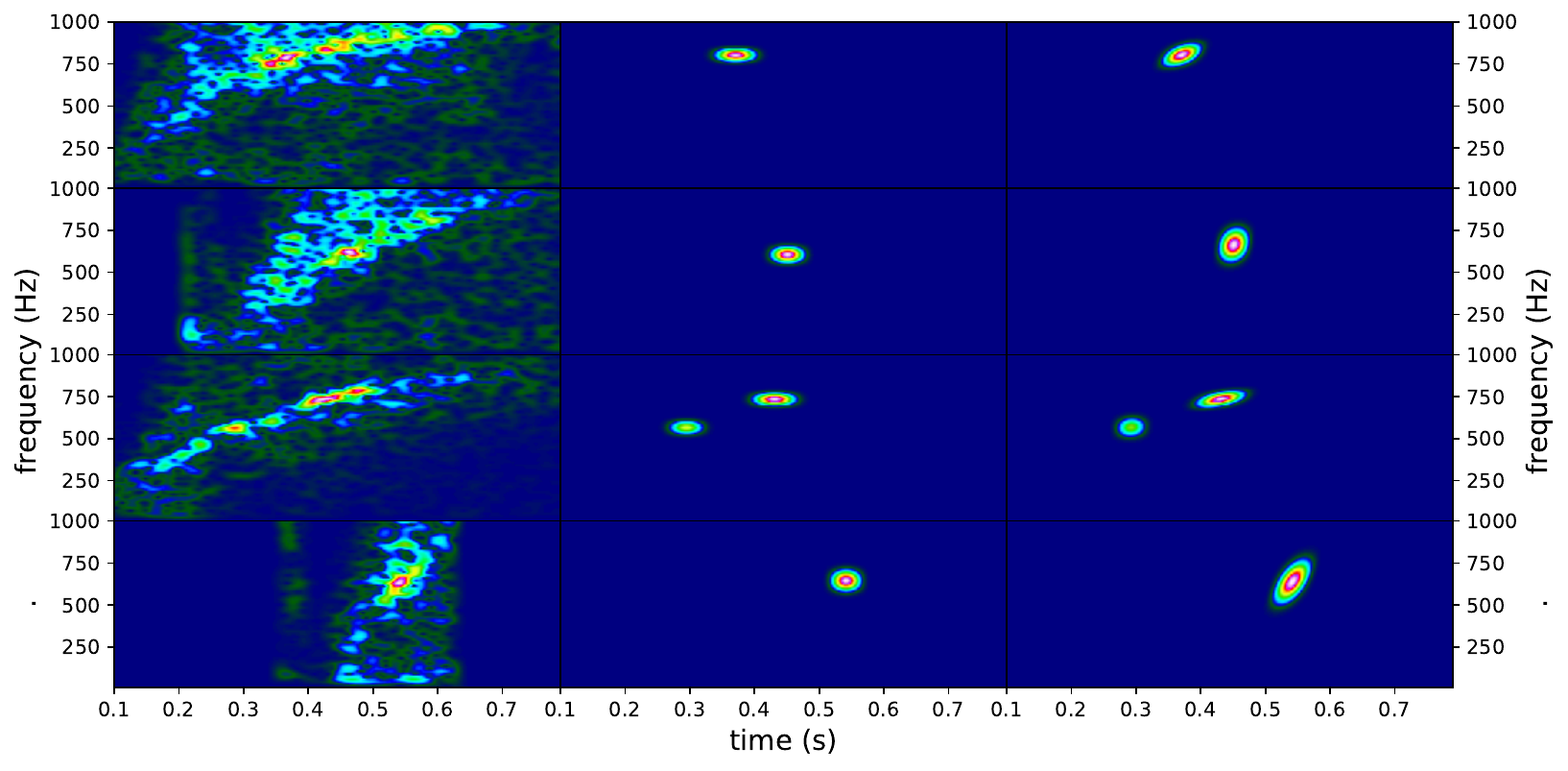}
\caption{ 
Same as Figure~\ref{fig:recons_snr60} and \ref{fig:recons_snr35}, however the signal-to-noise of the injected signal is 20. The majority of the injections are reconstructed with only a single wavelet. 
}
\label{fig:recons_snr20}
\end{figure*}

In Figures \ref{fig:recons_snr60},  \ref{fig:recons_snr35}, and \ref{fig:recons_snr20} we show example spectrograms of the waveform reconstructions at network SNR values of 60, 35, and 20 respectively. Spectrograms of the reconstructions are important, as spectrograms of CCSN signals show features that are directly related to the astrophysical properties of the CCSN. While overlap values are instructive to indicate overall goodness-of-fit, it is important to identify which features in reconstructions are informative about the source physics rather than simply stochastic features of the waveform. Therefore, looking at spectrogram reconstructions is key to understanding if the main important features are captured. 

At network SNR 60, the spectrogram reconstructions show that the dominant f/g-mode is well captured in the reconstructed waveforms. This is important as these reconstructions can be combined with universal relations for CCSN modes to measure the size of the PNS \cite[e.g.,][]{torres_19, sotani_21, powell_22}. For the s18 and z85 models, the mode is clearly better captured by the chirplet model. For the m39 model, both the sine Gaussian and the chirplet model provide good reconstructions, but more features are present in the sine Gaussian reconstruction. The y20 model has some loud gravitational-wave emission above the dominant mode. This additional feature makes the waveform more complex and therefore more difficult to reconstruct. The reconstructions at high SNR clearly also include some of this high frequency ``haze" above the dominant mode. 

In the lower SNR 35 reconstructions, the highest amplitude parts of the mode are still captured, which would still enable statements about the evolving size of the PNS. It is clear that at this SNR, the detected gravitational-wave signal for a CCSN may be a collection of disjointed blobs, instead of one continuous looking mode. More of the mode evolution is captured with the chirplet model for the s18, y20 and z85 signals. For the m39 model, a larger duration of the mode is captured with the sine Gaussian model, as it is constructed from more wavelets. This shows that even though the sine Gaussian wavelets often produce a slightly higher overlap, the chirplet model is better at capturing the CCSN signal features that are important for the astrophysical interpretation of the source. 

For the SNR 20 reconstructions, the reconstructions are only one or two wavelets, which capture only the highest amplitude part of the signal.
The mode amplitude is usually highest shortly after shock revival, making the shock revival time the most likely aspect of the gravitational-wave emission that will be reconstructed. This may still enable statements about the explosion mechanism even for the lowest possible SNR CCSN detections.

%%%%%%%%%%%%%%%%%%%%%%%%%%%%%%%%%%%%%%%%%%%%%%%%
\section{Conclusions}
\label{sec:conclusions}

In this work, we reconstruct four different CCSN waveform models with transdimensional Bayesian inference. We use two different types of wavelets: sine Gaussians and chirplets. We measure the overlap between the injected and reconstructed waveforms, and look at what CCSN signal features are captured in spectrograms of the reconstructed signals. 

We find that \texttt{tBilby} is capable of reconstructing up to 85\% of the CCSN signal features. As CCSN gravitational-wave signals have some unimportant stochastic elements, we find this reconstruction value is high enough to capture many of the features that are important for understanding the source properties, such as the size of the newly born PNS. Even for significantly lower reconstruction overlap values, most of the PNS mode is still visible in spectrograms of the reconstructed signal. 

In this study, we use a maximum frequency of 1024\,Hz, as it is more computationally expensive to go to higher frequency values. In reality, there have been multiple simulations of gravitational-wave emission from CCSNe when the frequency of the dominant mode goes well above 1000\,Hz \cite{pan_21, powell_23, powell_24b, choi_24}. So a larger frequency range would be needed for the analysis of a real CCSN detection. 

The CCSN waveforms included in this study mainly contain only one visible PNS mode. They were selected as the most simple first step for a CCSN waveform analysis. However, a significant number of CCSN waveforms from numerical simulations also contain additional features. Some examples are additional PNS modes, prompt convection or emission below a few 100\,Hz from the SASI. The next step to be ready for the analysis of a real CCSN detection will be to test this reconstruction method on these more complex CCSN waveforms. 

In the future, we could also use a larger variety of wavelets. We could use identical wavelets to other gravitational-wave reconstruction methods for a more direct code comparison. Previous work with \texttt{Bilby} has also used models consisting of multiple principal components to determine the CCSN explosion mechanism \cite{powell_24}, which is also a problem well suited to transdimensional inference where the number of principal components would be a free parameter.

%%%%%%%%%%%%%%%%%%%%%%%%%%%%%%%%%%%%%%%%%%%%%%%%%%%%%%%%
\section{Acknowledgments} The authors are supported by the Australian Research Council's (ARC) Centre of Excellence for Gravitational Wave Discovery (OzGrav) through project number CE230100016. JP and PDL acknowledge support from the ARC through LIEF Project LE260100008. PDL and NG are supported through ARC Discovery Project DP230103088.
This work was performed on the OzSTAR national facility at Swinburne University of Technology. The OzSTAR program receives funding in part from the Astronomy National Collaborative Research Infrastructure Strategy (NCRIS) allocation provided by the Australian Government, and from the Victorian Higher Education State Investment Fund (VHESIF) provided by the Victorian Government.



\bibliographystyle{apsrev}

\bibliography{example} % if your bibtex file is called example.bib

@ARTICLE{andresen_19,
       author = {{Andresen}, H. and {M{\"u}ller}, E. and {Janka}, H.-Th and {Summa}, A. and {Gill}, K. and {Zanolin}, M.},
        title = "{Gravitational waves from 3D core-collapse supernova models: The impact of moderate progenitor rotation}",
      journal = {\mnras},
         year = 2019,
        month = jun,
       volume = {486},
       number = {2},
        pages = {2238-2253},
          doi = {10.1093/mnras/stz990},
archivePrefix = {arXiv},
       eprint = {1810.07638},
 primaryClass = {astro-ph.HE},
       adsurl = {https://ui.adsabs.harvard.edu/abs/2019MNRAS.486.2238A}
}

@ARTICLE{blondin_03,
       author = {{Blondin}, John M. and {Mezzacappa}, Anthony and {DeMarino}, Christine},
        title = "{Stability of Standing Accretion Shocks, with an Eye toward Core-Collapse Supernovae}",
      journal = {\apj},
         year = 2003,
        month = feb,
       volume = {584},
       number = {2},
        pages = {971-980},
          doi = {10.1086/345812},
archivePrefix = {arXiv},
       eprint = {astro-ph/0210634},
 primaryClass = {astro-ph},
       adsurl = {https://ui.adsabs.harvard.edu/abs/2003ApJ...584..971B}
}

@ARTICLE{bruenn_94,
       author = {{Bruenn}, Stephen W. and {Mezzacappa}, Anthony},
        title = "{Prompt Convection in Core Collapse Supernovae}",
      journal = {\apjl},
         year = 1994,
        month = sep,
       volume = {433},
        pages = {L45},
          doi = {10.1086/187544},
       adsurl = {https://ui.adsabs.harvard.edu/abs/1994ApJ...433L..45B}
}

@ARTICLE{bruel_23,
       author = {{Bruel}, Tristan and {Bizouard}, Marie-Anne and {Obergaulinger}, Martin and {Maturana-Russel}, Patricio and {Torres-Forn{\'e}}, Alejandro and {Cerd{\'a}-Dur{\'a}n}, Pablo and {Christensen}, Nelson and {Font}, Jos{\'e} A. and {Meyer}, Renate},
        title = "{Inference of protoneutron star properties in core-collapse supernovae from a gravitational-wave detector network}",
      journal = {\prd},
         year = 2023,
        month = apr,
       volume = {107},
       number = {8},
          eid = {083029},
        pages = {083029},
          doi = {10.1103/PhysRevD.107.083029},
archivePrefix = {arXiv},
       eprint = {2301.10019},
 primaryClass = {astro-ph.HE},
       adsurl = {https://ui.adsabs.harvard.edu/abs/2023PhRvD.107h3029B}
}

@ARTICLE{burrows_21,
       author = {{Burrows}, A. and {Vartanyan}, D.},
        title = "{Core-collapse supernova explosion theory}",
      journal = {\nat},
         year = 2021,
        month = jan,
       volume = {589},
       number = {7840},
        pages = {29-39},
          doi = {10.1038/s41586-020-03059-w},
archivePrefix = {arXiv},
       eprint = {2009.14157},
 primaryClass = {astro-ph.SR},
       adsurl = {https://ui.adsabs.harvard.edu/abs/2021Natur.589...29B}
}

@ARTICLE{burrows_26,
       author = {{Burrows}, Adam and {Wang}, Tianshu and {Vartanyan}, David},
        title = "{Effects of Rotation on 3D Core-Collapse Supernova Models for Low-Mass Progenitors}",
      journal = {arXiv e-prints},
         year = 2026,
        month = jul,
          eid = {arXiv:2607.06664},
        pages = {arXiv:2607.06664},
          doi = {10.48550/arXiv.2607.06664},
archivePrefix = {arXiv},
       eprint = {2607.06664},
 primaryClass = {astro-ph.HE},
       adsurl = {https://ui.adsabs.harvard.edu/abs/2026arXiv260706664B}
}

@ARTICLE{choi_24,
       author = {{Choi}, Lyla and {Burrows}, Adam and {Vartanyan}, David},
        title = "{Gravitational-wave and Gravitational-wave Memory Signatures of Core-collapse Supernovae}",
      journal = {\apj},
         year = 2024,
        month = nov,
       volume = {975},
       number = {1},
          eid = {12},
        pages = {12},
          doi = {10.3847/1538-4357/ad74f8},
archivePrefix = {arXiv},
       eprint = {2408.01525},
 primaryClass = {astro-ph.HE},
       adsurl = {https://ui.adsabs.harvard.edu/abs/2024ApJ...975...12C}
}

@ARTICLE{cusinato_26,
       author = {{Cusinato}, M. and {Obergaulinger}, M. and {Aloy}, M. {\'A}.},
        title = "{Convection signatures in early-time gravitational waves from core-collapse supernovae}",
      journal = {\aap},
         year = 2026,
        month = jan,
       volume = {705},
          eid = {A179},
        pages = {A179},
          doi = {10.1051/0004-6361/202556543},
archivePrefix = {arXiv},
       eprint = {2507.16903},
 primaryClass = {astro-ph.HE},
       adsurl = {https://ui.adsabs.harvard.edu/abs/2026A&A...705A.179C}
}

@ARTICLE{cornish_15,
       author = {{Cornish}, Neil J. and {Littenberg}, Tyson B.},
        title = "{Bayeswave: Bayesian inference for gravitational wave bursts and instrument glitches}",
      journal = {Classical and Quantum Gravity},
         year = 2015,
        month = jul,
       volume = {32},
       number = {13},
          eid = {135012},
        pages = {135012},
          doi = {10.1088/0264-9381/32/13/135012},
archivePrefix = {arXiv},
       eprint = {1410.3835},
 primaryClass = {gr-qc},
       adsurl = {https://ui.adsabs.harvard.edu/abs/2015CQGra..32m5012C}
}

@ARTICLE{drago_21,
       author = {{Drago}, Marco and {Klimenko}, Sergey and {Lazzaro}, Claudia and {Milotti}, Edoardo and {Mitselmakher}, Guenakh and {Necula}, Valentin and {O'Brian}, Brendan and {Prodi}, Giovanni Andrea and {Salemi}, Francesco and {Szczepanczyk}, Marek and {Tiwari}, Shubhanshu and {Tiwari}, Vaibhav and {Gayathri}, V. and {Vedovato}, Gabriele and {Yakushin}, Igor},
        title = "{coherent WaveBurst, a pipeline for unmodeled gravitational-wave data analysis}",
      journal = {SoftwareX},
         year = 2021,
        month = jun,
       volume = {14},
          eid = {100678},
        pages = {100678},
          doi = {10.1016/j.softx.2021.100678},
archivePrefix = {arXiv},
       eprint = {2006.12604},
 primaryClass = {gr-qc},
       adsurl = {https://ui.adsabs.harvard.edu/abs/2021SoftX..1400678D}
}

@ARTICLE{speagle_20,
       author = {{Speagle}, Joshua S.},
        title = "{DYNESTY: a dynamic nested sampling package for estimating Bayesian posteriors and evidences}",
      journal = {\mnras},
         year = 2020,
        month = apr,
       volume = {493},
       number = {3},
        pages = {3132-3158},
          doi = {10.1093/mnras/staa278},
archivePrefix = {arXiv},
       eprint = {1904.02180},
 primaryClass = {astro-ph.IM},
       adsurl = {https://ui.adsabs.harvard.edu/abs/2020MNRAS.493.3132S}
}

@ARTICLE{edwards_21,
       author = {{Edwards}, Matthew C.},
        title = "{Classifying the equation of state from rotating core collapse gravitational waves with deep learning}",
      journal = {\prd},
         year = 2021,
        month = jan,
       volume = {103},
       number = {2},
          eid = {024025},
        pages = {024025},
          doi = {10.1103/PhysRevD.103.024025},
archivePrefix = {arXiv},
       eprint = {2009.07367},
 primaryClass = {astro-ph.IM},
       adsurl = {https://ui.adsabs.harvard.edu/abs/2021PhRvD.103b4025E}
}

@ARTICLE{foglizzo_24,
       author = {{Foglizzo}, T.},
        title = "{Analytic insight into the physics of the standing accretion shock instability: I. Shock instability in a non-rotating stellar core}",
      journal = {\aap},
         year = 2024,
        month = dec,
       volume = {692},
          eid = {A196},
        pages = {A196},
          doi = {10.1051/0004-6361/202452300},
archivePrefix = {arXiv},
       eprint = {2409.12725},
 primaryClass = {astro-ph.HE},
       adsurl = {https://ui.adsabs.harvard.edu/abs/2024A&A...692A.196F}
}

@ARTICLE{gwtc-5,
       author = {{The LIGO Scientific Collaboration} and {the Virgo Collaboration} and {the KAGRA Collaboration}},
        title = "{GWTC-5.0: Observations from the Second Part of the Fourth LIGO-Virgo-KAGRA Observing Run and Updates to the Gravitational-Wave Transient Catalog}",
      journal = {arXiv e-prints},
         year = 2026,
        month = may,
          eid = {arXiv:2605.27225},
        pages = {arXiv:2605.27225},
          doi = {10.48550/arXiv.2605.27225},
archivePrefix = {arXiv},
       eprint = {2605.27225},
 primaryClass = {gr-qc},
       adsurl = {https://ui.adsabs.harvard.edu/abs/2026arXiv260527225T}
}

@ARTICLE{jakobus_23,
       author = {{Jakobus}, Pia and {M{\"u}ller}, Bernhard and {Heger}, Alexander and {Zha}, Shuai and {Powell}, Jade and {Motornenko}, Anton and {Steinheimer}, Jan and {St{\"o}cker}, Horst},
        title = "{Gravitational Waves from a Core g Mode in Supernovae as Probes of the High-Density Equation of State}",
      journal = {\prl},
         year = 2023,
        month = nov,
       volume = {131},
       number = {19},
          eid = {191201},
        pages = {191201},
          doi = {10.1103/PhysRevLett.131.191201},
archivePrefix = {arXiv},
       eprint = {2301.06515},
 primaryClass = {astro-ph.HE},
       adsurl = {https://ui.adsabs.harvard.edu/abs/2023PhRvL.131s1201J}
}

@INCOLLECTION{janka_17,
       author = {{Janka}, Hans-Thomas},
        title = "{Neutrino-Driven Explosions}",
    booktitle = {Handbook of Supernovae},
         year = 2017,
       editor = {{Alsabti}, Athem W. and {Murdin}, Paul},
        pages = {1095},
          doi = {10.1007/978-3-319-21846-5_109},
       adsurl = {https://ui.adsabs.harvard.edu/abs/2017hsn..book.1095J}
}

@INPROCEEDINGS{jerkstrand_26,
       author = {{Jerkstrand}, Anders and {Milisavljevic}, Dan and {M{\"u}ller}, Bernhard},
        title = "{Core-collapse supernovae}",
    booktitle = {Encyclopedia of Astrophysics, Volume 2},
         year = 2026,
       volume = {2},
        month = jan,
        pages = {639-668},
          doi = {10.1016/B978-0-443-21439-4.00090-0},
archivePrefix = {arXiv},
       eprint = {2503.01321},
 primaryClass = {astro-ph.HE},
       adsurl = {https://ui.adsabs.harvard.edu/abs/2026enap....2..639J}
}

@ARTICLE{kagra_19,
       author = {{Kagra Collaboration} and {Akutsu}, T. and {Ando}, M. and {Arai}, K. and {Arai}, Y. and {Araki}, S. and {Araya}, A. and {Aritomi}, N. and {Asada}, H. and {Aso}, Y. and {Atsuta}, S. and others},
        title = "{KAGRA: 2.5 generation interferometric gravitational wave detector}",
      journal = {Nature Astronomy},
         year = 2019,
        month = jan,
       volume = {3},
        pages = {35-40},
          doi = {10.1038/s41550-018-0658-y},
archivePrefix = {arXiv},
       eprint = {1811.08079},
 primaryClass = {gr-qc},
       adsurl = {https://ui.adsabs.harvard.edu/abs/2019NatAs...3...35K}
}

@ARTICLE{kuroda_17,
       author = {{Kuroda}, Takami and {Kotake}, Kei and {Hayama}, Kazuhiro and {Takiwaki}, Tomoya},
        title = "{Correlated Signatures of Gravitational-wave and Neutrino Emission in Three-dimensional General-relativistic Core-collapse Supernova Simulations}",
      journal = {\apj},
         year = 2017,
        month = dec,
       volume = {851},
       number = {1},
          eid = {62},
        pages = {62},
          doi = {10.3847/1538-4357/aa988d},
archivePrefix = {arXiv},
       eprint = {1708.05252},
 primaryClass = {astro-ph.HE},
       adsurl = {https://ui.adsabs.harvard.edu/abs/2017ApJ...851...62K}
}

@ARTICLE{kuroda_20,
       author = {{Kuroda}, Takami and {Arcones}, Almudena and {Takiwaki}, Tomoya and {Kotake}, Kei},
        title = "{Magnetorotational Explosion of a Massive Star Supported by Neutrino Heating in General Relativistic Three-dimensional Simulations}",
      journal = {\apj},
         year = 2020,
        month = jun,
       volume = {896},
       number = {2},
          eid = {102},
        pages = {102},
          doi = {10.3847/1538-4357/ab9308},
archivePrefix = {arXiv},
       eprint = {2003.02004},
 primaryClass = {astro-ph.HE},
       adsurl = {https://ui.adsabs.harvard.edu/abs/2020ApJ...896..102K}
}

@ARTICLE{lee_25,
       author = {{Lee}, Yi Shuen C. and {Szczepa{\'n}czyk}, Marek J. and {Mishra}, Tanmaya and {Millhouse}, Margaret and {Melatos}, Andrew},
        title = "{Dedicated-frequency analysis of gravitational-wave bursts from core-collapse supernovae with minimal assumptions}",
      journal = {\prd},
         year = 2025,
        month = oct,
       volume = {112},
       number = {8},
          eid = {082006},
        pages = {082006},
          doi = {10.1103/kg3l-dtxc},
archivePrefix = {arXiv},
       eprint = {2510.01614},
 primaryClass = {astro-ph.HE},
       adsurl = {https://ui.adsabs.harvard.edu/abs/2025PhRvD.112h2006L}
}

@online{ligo_noise_curve,
  author = {LIGO Scientific Collaboration, Virgo Collaboration, KAGRA Collabroation},
  title = {Virgo and KAGRA noise curves for use in simulations},
  year = 2022,
  url = {https://dcc.ligo.org/LIGO-T2200043},
  urldate = {15-12-2022}
}

@ARTICLE{ligo_15,
       author = {{LIGO Scientific Collaboration} and {Aasi}, J. and {Abbott}, B.~P. and {Abbott}, R. and {Abbott}, T. and {Abernathy}, M.~R. and {Ackley}, K. and {Adams}, C. and {Adams}, T. and {Addesso}, P. and {Adhikari}, R.~X. and {Adya}, V. and {Affeldt}, C. and {Aggarwal}, N. and others},
        title = "{Advanced LIGO}",
      journal = {Classical and Quantum Gravity},
         year = 2015,
        month = apr,
       volume = {32},
       number = {7},
          eid = {074001},
        pages = {074001},
          doi = {10.1088/0264-9381/32/7/074001},
archivePrefix = {arXiv},
       eprint = {1411.4547},
 primaryClass = {gr-qc},
       adsurl = {https://ui.adsabs.harvard.edu/abs/2015CQGra..32g4001L}
}

@ARTICLE{mezzacappa_24,
       author = {{Mezzacappa}, Anthony and {Zanolin}, Michele},
        title = "{Gravitational Waves from Neutrino-Driven Core Collapse Supernovae: Predictions, Detection, and Parameter Estimation}",
      journal = {arXiv e-prints},
         year = 2024,
        month = jan,
          eid = {arXiv:2401.11635},
        pages = {arXiv:2401.11635},
          doi = {10.48550/arXiv.2401.11635},
archivePrefix = {arXiv},
       eprint = {2401.11635},
 primaryClass = {astro-ph.HE},
       adsurl = {https://ui.adsabs.harvard.edu/abs/2024arXiv240111635M}
}

@ARTICLE{mueller_24,
       author = {{M{\"u}ller}, Bernhard},
        title = "{Supernova Simulations}",
      journal = {arXiv e-prints},
         year = 2024,
        month = mar,
          eid = {arXiv:2403.18952},
        pages = {arXiv:2403.18952},
          doi = {10.48550/arXiv.2403.18952},
archivePrefix = {arXiv},
       eprint = {2403.18952},
 primaryClass = {astro-ph.HE},
       adsurl = {https://ui.adsabs.harvard.edu/abs/2024arXiv240318952M}
}

@ARTICLE{millhouse_18,
       author = {{Millhouse}, Margaret and {Cornish}, Neil J. and {Littenberg}, Tyson},
        title = "{Bayesian reconstruction of gravitational wave bursts using chirplets}",
      journal = {\prd},
         year = 2018,
        month = may,
       volume = {97},
       number = {10},
          eid = {104057},
        pages = {104057},
          doi = {10.1103/PhysRevD.97.104057},
archivePrefix = {arXiv},
       eprint = {1804.03239},
 primaryClass = {gr-qc},
       adsurl = {https://ui.adsabs.harvard.edu/abs/2018PhRvD..97j4057M}
}

@ARTICLE{oconnor_18,
       author = {{O'Connor}, Evan P. and {Couch}, Sean M.},
        title = "{Exploring Fundamentally Three-dimensional Phenomena in High-fidelity Simulations of Core-collapse Supernovae}",
      journal = {\apj},
         year = 2018,
        month = oct,
       volume = {865},
       number = {2},
          eid = {81},
        pages = {81},
          doi = {10.3847/1538-4357/aadcf7},
archivePrefix = {arXiv},
       eprint = {1807.07579},
 primaryClass = {astro-ph.HE},
       adsurl = {https://ui.adsabs.harvard.edu/abs/2018ApJ...865...81O}
}

@ARTICLE{pastor_24,
       author = {{Pastor-Marcos}, Carlos and {Cerd{\'a}-Dur{\'a}n}, Pablo and {Walker}, Daniel and {Torres-Forn{\'e}}, Alejandro and {Abdikamalov}, Ernazar and {Richers}, Sherwood and {Font}, Jos{\'e} A.},
        title = "{Bayesian inference from gravitational waves in fast-rotating, core-collapse supernovae}",
      journal = {\prd},
         year = 2024,
        month = mar,
       volume = {109},
       number = {6},
          eid = {063028},
        pages = {063028},
          doi = {10.1103/PhysRevD.109.063028},
archivePrefix = {arXiv},
       eprint = {2308.03456},
 primaryClass = {astro-ph.HE},
       adsurl = {https://ui.adsabs.harvard.edu/abs/2024PhRvD.109f3028P}
}

@ARTICLE{pan_21,
       author = {{Pan}, Kuo-Chuan and {Liebend{\"o}rfer}, Matthias and {Couch}, Sean M. and {Thielemann}, Friedrich-Karl},
        title = "{Stellar Mass Black Hole Formation and Multimessenger Signals from Three-dimensional Rotating Core-collapse Supernova Simulations}",
      journal = {\apj},
         year = 2021,
        month = jun,
       volume = {914},
       number = {2},
          eid = {140},
        pages = {140},
          doi = {10.3847/1538-4357/abfb05},
archivePrefix = {arXiv},
       eprint = {2010.02453},
 primaryClass = {astro-ph.HE},
       adsurl = {https://ui.adsabs.harvard.edu/abs/2021ApJ...914..140P}
}

@ARTICLE{powell_19,
       author = {{Powell}, Jade and {M{\"u}ller}, Bernhard},
        title = "{Gravitational wave emission from 3D explosion models of core-collapse supernovae with low and normal explosion energies}",
      journal = {\mnras},
         year = 2019,
        month = jul,
       volume = {487},
       number = {1},
        pages = {1178-1190},
          doi = {10.1093/mnras/stz1304},
archivePrefix = {arXiv},
       eprint = {1812.05738},
 primaryClass = {astro-ph.HE},
       adsurl = {https://ui.adsabs.harvard.edu/abs/2019MNRAS.487.1178P}
}

@ARTICLE{powell_20,
       author = {{Powell}, Jade and {M{\"u}ller}, Bernhard},
        title = "{Three-dimensional core-collapse supernova simulations of massive and rotating progenitors}",
      journal = {\mnras},
         year = 2020,
        month = jun,
       volume = {494},
       number = {4},
        pages = {4665-4675},
          doi = {10.1093/mnras/staa1048},
archivePrefix = {arXiv},
       eprint = {2002.10115},
 primaryClass = {astro-ph.HE},
       adsurl = {https://ui.adsabs.harvard.edu/abs/2020MNRAS.494.4665P}
}

@ARTICLE{powell_21,
       author = {{Powell}, Jade and {M{\"u}ller}, Bernhard and {Heger}, Alexander},
        title = "{The final core collapse of pulsational pair instability supernovae}",
      journal = {\mnras},
         year = 2021,
        month = may,
       volume = {503},
       number = {2},
        pages = {2108-2122},
          doi = {10.1093/mnras/stab614},
archivePrefix = {arXiv},
       eprint = {2101.06889},
 primaryClass = {astro-ph.HE},
       adsurl = {https://ui.adsabs.harvard.edu/abs/2021MNRAS.503.2108P}
}

@ARTICLE{powell_22,
       author = {{Powell}, Jade and {M{\"u}ller}, Bernhard},
        title = "{Inferring astrophysical parameters of core-collapse supernovae from their gravitational-wave emission}",
      journal = {\prd},
         year = 2022,
        month = mar,
       volume = {105},
       number = {6},
          eid = {063018},
        pages = {063018},
          doi = {10.1103/PhysRevD.105.063018},
archivePrefix = {arXiv},
       eprint = {2201.01397},
 primaryClass = {astro-ph.HE},
       adsurl = {https://ui.adsabs.harvard.edu/abs/2022PhRvD.105f3018P}
}

@ARTICLE{powell_23,
       author = {{Powell}, Jade and {M{\"u}ller}, Bernhard and {Aguilera-Dena}, David R. and {Langer}, Norbert},
        title = "{Three dimensional magnetorotational core-collapse supernova explosions of a 39 solar mass progenitor star}",
      journal = {\mnras},
         year = 2023,
        month = jul,
       volume = {522},
       number = {4},
        pages = {6070-6086},
          doi = {10.1093/mnras/stad1292},
archivePrefix = {arXiv},
       eprint = {2212.00200},
 primaryClass = {astro-ph.HE},
       adsurl = {https://ui.adsabs.harvard.edu/abs/2023MNRAS.522.6070P}
}

@ARTICLE{powell_24,
       author = {{Powell}, Jade and {Iess}, Alberto and {Llorens-Monteagudo}, Miquel and {Obergaulinger}, Martin and {M{\"u}ller}, Bernhard and {Torres-Forn{\'e}}, Alejandro and {Cuoco}, Elena and {Font}, Jos{\'e} A.},
        title = "{Determining the core-collapse supernova explosion mechanism with current and future gravitational-wave observatories}",
      journal = {\prd},
         year = 2024,
        month = mar,
       volume = {109},
       number = {6},
          eid = {063019},
        pages = {063019},
          doi = {10.1103/PhysRevD.109.063019},
archivePrefix = {arXiv},
       eprint = {2311.18221},
 primaryClass = {astro-ph.HE},
       adsurl = {https://ui.adsabs.harvard.edu/abs/2024PhRvD.109f3019P}
}

@ARTICLE{powell_24b,
       author = {{Powell}, Jade and {M{\"u}ller}, Bernhard},
        title = "{The gravitational-wave emission from the explosion of a 15 solar mass star with rotation and magnetic fields}",
      journal = {\mnras},
         year = 2024,
        month = aug,
       volume = {532},
       number = {4},
        pages = {4326-4339},
          doi = {10.1093/mnras/stae1731},
archivePrefix = {arXiv},
       eprint = {2406.09691},
 primaryClass = {astro-ph.HE},
       adsurl = {https://ui.adsabs.harvard.edu/abs/2024MNRAS.532.4326P}
}

@ARTICLE{powell_25,
       author = {{Powell}, Jade and {Lasky}, Paul D.},
        title = "{The Dawes Review 12: Gravitational-wave burst astrophysics}",
      journal = {PASA},
         year = 2025,
        month = mar,
       volume = {42},
          eid = {e030},
        pages = {e030},
          doi = {10.1017/pasa.2025.10},
archivePrefix = {arXiv},
       eprint = {2410.12105},
 primaryClass = {astro-ph.HE},
       adsurl = {https://ui.adsabs.harvard.edu/abs/2025PASA...42...30P}
}

@ARTICLE{raza_22,
       author = {{Raza}, Nayyer and {McIver}, Jess and {D{\'a}lya}, Gergely and {Raffai}, Peter},
        title = "{Prospects for reconstructing the gravitational-wave signals from core-collapse supernovae with Advanced LIGO-Virgo and the BayesWave algorithm}",
      journal = {\prd},
         year = 2022,
        month = sep,
       volume = {106},
       number = {6},
          eid = {063014},
        pages = {063014},
          doi = {10.1103/PhysRevD.106.063014},
archivePrefix = {arXiv},
       eprint = {2203.08960},
 primaryClass = {astro-ph.HE},
       adsurl = {https://ui.adsabs.harvard.edu/abs/2022PhRvD.106f3014R}
}

@ARTICLE{radice_19,
       author = {{Radice}, David and {Morozova}, Viktoriya and {Burrows}, Adam and {Vartanyan}, David and {Nagakura}, Hiroki},
        title = "{Characterizing the Gravitational Wave Signal from Core-collapse Supernovae}",
      journal = {\apjl},
         year = 2019,
        month = may,
       volume = {876},
       number = {1},
          eid = {L9},
        pages = {L9},
          doi = {10.3847/2041-8213/ab191a},
archivePrefix = {arXiv},
       eprint = {1812.07703},
 primaryClass = {astro-ph.HE},
       adsurl = {https://ui.adsabs.harvard.edu/abs/2019ApJ...876L...9R}
}

@ARTICLE{sotani_21,
       author = {{Sotani}, Hajime and {Takiwaki}, Tomoya and {Togashi}, Hajime},
        title = "{Universal relation for supernova gravitational waves}",
      journal = {\prd},
         year = 2021,
        month = dec,
       volume = {104},
       number = {12},
          eid = {123009},
        pages = {123009},
          doi = {10.1103/PhysRevD.104.123009},
archivePrefix = {arXiv},
       eprint = {2110.03131},
 primaryClass = {astro-ph.HE},
       adsurl = {https://ui.adsabs.harvard.edu/abs/2021PhRvD.104l3009S}
}

@ARTICLE{sn2023ixf_25,
       author = {{Abac}, A.~G. and {Abbott}, R. and {Abouelfettouh}, I. and {Acernese}, F. and {Ackley}, K. and {Adhicary}, S. and others},
        title = "{Search for Gravitational Waves Emitted from SN 2023ixf}",
      journal = {\apj},
         year = 2025,
        month = jun,
       volume = {985},
       number = {2},
          eid = {183},
        pages = {183},
          doi = {10.3847/1538-4357/adc681},
archivePrefix = {arXiv},
       eprint = {2410.16565},
 primaryClass = {astro-ph.HE},
       adsurl = {https://ui.adsabs.harvard.edu/abs/2025ApJ...985..183A}
}

@ARTICLE{szczepanczyk_21,
       author = {{Szczepa{\'n}czyk}, Marek J. and {Antelis}, Javier M. and {Benjamin}, Michael and {Cavagli{\`a}}, Marco and {Gondek-Rosi{\'n}ska}, Dorota and {Hansen}, Travis and {Klimenko}, Sergey and {Morales}, Manuel D. and {Moreno}, Claudia and {Mukherjee}, Soma and {Nurbek}, Gaukhar and {Powell}, Jade and {Singh}, Neha and {Sitmukhambetov}, Satzhan and {Szewczyk}, Pawe{\l} and {Valdez}, Oscar and {Vedovato}, Gabriele and {Westhouse}, Jonathan and {Zanolin}, Michele and {Zheng}, Yanyan},
        title = "{Detecting and reconstructing gravitational waves from the next galactic core-collapse supernova in the advanced detector era}",
      journal = {\prd},
         year = 2021,
        month = nov,
       volume = {104},
       number = {10},
          eid = {102002},
        pages = {102002},
          doi = {10.1103/PhysRevD.104.102002},
archivePrefix = {arXiv},
       eprint = {2104.06462},
 primaryClass = {astro-ph.HE},
       adsurl = {https://ui.adsabs.harvard.edu/abs/2021PhRvD.104j2002S}
}

@ARTICLE{szczepanczyk_24,
       author = {{Szczepa{\'n}czyk}, Marek J. and {Zheng}, Yanyan and {Antelis}, Javier M. and {Benjamin}, Michael and {Bizouard}, Marie-Anne and {Casallas-Lagos}, Alejandro and {Cerd{\'a}-Dur{\'a}n}, Pablo and {Davis}, Derek and {Gondek-Rosi{\'n}ska}, Dorota and {Klimenko}, Sergey and {Moreno}, Claudia and {Obergaulinger}, Martin and {Powell}, Jade and {Ramirez}, Dymetris and {Ratto}, Brad and {Richardson}, Colter and {Rijal}, Abhinav and {Stuver}, Amber L. and {Szewczyk}, Pawe{\l} and {Vedovato}, Gabriele and {Zanolin}, Michele and {Bartos}, Imre and {Bhaumik}, Shubhagata and {Bulik}, Tomasz and {Drago}, Marco and {Font}, Jos{\'e} A. and {De Colle}, Fabio and {Garc{\'\i}a-Bellido}, Juan and {Gayathri}, V. and {Hughey}, Brennan and {Mitselmakher}, Guenakh and {Mishra}, Tanmaya and {Mukherjee}, Soma and {Nguyen}, Quynh Lan and {Chan}, Man Leong and {Di Palma}, Irene and {Piotrzkowski}, Brandon J. and {Singh}, Neha},
        title = "{Optically targeted search for gravitational waves emitted by core-collapse supernovae during the third observing run of Advanced LIGO and Advanced Virgo}",
      journal = {\prd},
         year = 2024,
        month = aug,
       volume = {110},
       number = {4},
          eid = {042007},
        pages = {042007},
          doi = {10.1103/PhysRevD.110.042007},
archivePrefix = {arXiv},
       eprint = {2305.16146},
 primaryClass = {astro-ph.HE},
       adsurl = {https://ui.adsabs.harvard.edu/abs/2024PhRvD.110d2007S}
}

@ARTICLE{torres_19,
       author = {{Torres-Forn{\'e}}, Alejandro and {Cerd{\'a}-Dur{\'a}n}, Pablo and {Obergaulinger}, Martin and {M{\"u}ller}, Bernhard and {Font}, Jos{\'e} A.},
        title = "{Universal Relations for Gravitational-Wave Asteroseismology of Protoneutron Stars}",
      journal = {\prl},
         year = 2019,
        month = aug,
       volume = {123},
       number = {5},
          eid = {051102},
        pages = {051102},
          doi = {10.1103/PhysRevLett.123.051102},
archivePrefix = {arXiv},
       eprint = {1902.10048},
 primaryClass = {gr-qc},
       adsurl = {https://ui.adsabs.harvard.edu/abs/2019PhRvL.123e1102T}
}

@ARTICLE{tong_25,
       author = {{Tong}, Hui and {Guttman}, Nir and {Clarke}, Teagan A. and {Lasky}, Paul D. and {Thrane}, Eric and {Payne}, Ethan and {Nathan}, Rowina and {Farr}, Ben and {Fishbach}, Maya and {Ashton}, Gregory and {Marco}, Valentina Di},
        title = "{Transdimensional Inference for Gravitational-wave Astronomy with Bilby}",
      journal = {\apjs},
         year = 2025,
        month = feb,
       volume = {276},
       number = {2},
          eid = {50},
        pages = {50},
          doi = {10.3847/1538-4365/ad9deb},
archivePrefix = {arXiv},
       eprint = {2404.04460},
 primaryClass = {gr-qc},
       adsurl = {https://ui.adsabs.harvard.edu/abs/2025ApJS..276...50T}
}

@ARTICLE{walk_23,
       author = {{Walk}, Laurie and {Foglizzo}, Thierry and {Tamborra}, Irene},
        title = "{Standing accretion shock instability in the collapse of a rotating stellar core}",
      journal = {\prd},
         year = 2023,
        month = mar,
       volume = {107},
       number = {6},
          eid = {063014},
        pages = {063014},
          doi = {10.1103/PhysRevD.107.063014},
archivePrefix = {arXiv},
       eprint = {2212.07467},
 primaryClass = {astro-ph.HE},
       adsurl = {https://ui.adsabs.harvard.edu/abs/2023PhRvD.107f3014W}
}

@ARTICLE{vartanyan_20,
       author = {{Vartanyan}, David and {Burrows}, Adam},
        title = "{Gravitational Waves from Neutrino Emission Asymmetries in Core-collapse Supernovae}",
      journal = {\apj},
         year = 2020,
        month = oct,
       volume = {901},
       number = {2},
          eid = {108},
        pages = {108},
          doi = {10.3847/1538-4357/abafac},
archivePrefix = {arXiv},
       eprint = {2007.07261},
 primaryClass = {astro-ph.HE},
       adsurl = {https://ui.adsabs.harvard.edu/abs/2020ApJ...901..108V}
}

@ARTICLE{virgo_15,
       author = {{Acernese}, F. and {Agathos}, M. and {Agatsuma}, K. and {Aisa}, D. and {Allemandou}, N. and {Allocca}, A. and {Amarni}, J. and {Astone}, P. and {Balestri}, G. and {Ballardin}, G. and {Barone}, F. and {Baronick}, J.-P. and {Barsuglia}, M. and {Basti}, A. and others},
        title = "{Advanced Virgo: a second-generation interferometric gravitational wave detector}",
      journal = {Classical and Quantum Gravity},
         year = 2015,
        month = jan,
       volume = {32},
       number = {2},
          eid = {024001},
        pages = {024001},
          doi = {10.1088/0264-9381/32/2/024001},
archivePrefix = {arXiv},
       eprint = {1408.3978},
 primaryClass = {gr-qc},
       adsurl = {https://ui.adsabs.harvard.edu/abs/2015CQGra..32b4001A}
}

@article{bilby_paper,
    author = "Ashton, Gregory and others",
    title = "{BILBY: A user-friendly Bayesian inference library for gravitational-wave astronomy}",
    eprint = "1811.02042",
    archivePrefix = "arXiv",
    primaryClass = "astro-ph.IM",
    doi = "10.3847/1538-4365/ab06fc",
    journal = "Astrophys. J. Suppl.",
    volume = "241",
    number = "2",
    pages = "27",
    year = "2019"
}

%\begin{thebibliography}{99}
%\end{thebibliography}

%%%%%%%%%%%%%%%%%%%%%%%%%%%%%%%%%%%%%%%%%%%%%%%%%%

%%%%%%%%%%%%%%%%% APPENDICES %%%%%%%%%%%%%%%%%%%%%

% \appendix

% \section{Some extra material}

% If you want to present additional material which would interrupt the flow of the main paper,
% it can be placed in an Appendix which appears after the list of references.

%%%%%%%%%%%%%%%%%%%%%%%%%%%%%%%%%%%%%%%%%%%%%%%%%%

% Don't change these lines

\label{lastpage}
\end{document}